\documentclass[3p,twocolumn]{elsarticle}
\usepackage{lineno,hyperref,amsmath,amssymb}
\newcommand{ \eq}[1]{Eq.~(\ref{eq:#1})}
\modulolinenumbers[5]
\biboptions{sort&compress}

\makeatletter
\def\ps@pprintTitle{%
  \let\@oddhead\@empty
  \let\@evenhead\@empty
  \let\@oddfoot\@empty
  \let\@evenfoot\@oddfoot}
\makeatother

\begin{document}
\begin{frontmatter}
\title{{\bf Hadronic vacuum polarization contributions \\ to the muon {\boldmath $g$-2\unboldmath} in the space-like region}~\tnoteref{t1}}\tnotetext[t1]{This article is dedicated to the memory of Alberto Sirlin, a brilliant physicist and a superb mentor.}

\author[mymainaddress,mysecondaryaddress]{Elisa Balzani}
\author[mysecondaryaddress,mythirdaddress]{Stefano Laporta}
\author[mysecondaryaddress]{Massimo Passera}

\address[mymainaddress]{Dipartimento di Fisica e Astronomia `G.~Galilei', Universit\`a di Padova, Italy}
\address[mysecondaryaddress]{Istituto Nazionale di Fisica Nucleare, Sezione di Padova, Padova, Italy}
\address[mythirdaddress]{Dipartimento di Fisica, Universit\`a di Bologna e Istituto Nazionale di Fisica Nucleare, Sezione di Bologna, Bologna, Italy}

\begin{abstract}
We present simple analytic expressions to compute the hadronic vacuum polarization contribution to the muon $g$-2 in the space-like region up to next-to-next-to-leading order. These results can be employed in lattice QCD calculations of this contribution as well as in space-like determinations based on scattering data, like that expected from the proposed MUonE experiment at CERN.
\end{abstract}
%\begin{keyword}
%\end{keyword}
\end{frontmatter}

%%%%%%%%%%%%%%%%%%%%%%%%%%%%%%%%%%%%%%%%%%%%%%%%%%%%%%%%%%%%
\section{Introduction}
\label{Introduction}

The Muon $g$-2 (E989) experiment at Fermilab has recently presented its first measurement of the muon magnetic moment anomaly, $a_{\mu} = (g_{\mu}-2)/2$~\cite{Muong-2:2021ojo,Muong-2:2021vma,Muong-2:2021ovs,Muong-2:2021xzz}, confirming the earlier results of the E821 experiment at Brookhaven~\cite{Muong-2:2006rrc}. The E989 experiment is expected to reach a sensitivity four-times better than the E821 one. In addition, a new low-energy approach to measuring the muon $g$-2 is being developed by the E34 collaboration at J-PARC~\cite{Abe:2019thb}.

The present muon $g$-2 experimental average shows an intriguing $4.2 \sigma$ discrepancy with the value of the Standard Model (SM) $a_{\mu}$ prediction quoted by the Muon $g$-2 Theory Initiative~\cite{Aoyama:2020ynm}. If confirmed with high significance, this discrepancy would be indirect evidence for new physics beyond the SM.

The main uncertainty of the muon $g$-2 SM prediction originates from its hadronic vacuum polarization (HVP) contribution, $a_{\mu}^{\rm HVP}$, which cannot be reliably calculated perturbatively in QCD and relies on experimental data as input to dispersion relations. Indeed, this contribution has been traditionally computed via a dispersive, or time-like, integral using hadronic production cross sections in low-energy electron-positron annihilation. The present time-like calculation of $a_{\mu}^{\rm HVP}$ includes the leading-order (LO), next-to-leading-order (NLO) and next-to-next-to-leading-order (NNLO) terms~\cite{Jegerlehner:2017gek,Davier:2017zfy,Keshavarzi:2018mgv,Colangelo:2018mtw,Hoferichter:2019mqg,Davier:2019can,Keshavarzi:2019abf,Hoid:2020xjs,Kurz:2014wya}. The NNLO term is comparable to the final uncertainty of the $a_{\mu}$ measurement expected from the Muon $g$-2 experiment at Fermilab.

An alternative determination of $a_{\mu}^{\rm HVP}$ can be provided by lattice QCD~\cite{Chakraborty:2017tqp,Borsanyi:2017zdw,Blum:2018mom,Giusti:2019xct,Shintani:2019wai,FermilabLattice:2019ugu,Gerardin:2019rua,Aubin:2019usy,Giusti:2019hkz,Chakraborty:2018iyb}. Significant progress has been made in the last few years in first-principles lattice QCD calculations of its LO part, $a_{\mu}^{\rm HVP}({\rm LO})$, although the precision of these results is, in general, not yet competitive with that of the time-like determinations based on experimental data. Recently, the BMW collaboration presented the first lattice QCD calculation of $a_{\mu}^{\rm HVP}({\rm LO})$ with an impressive sub-percent (0.8\%) relative accuracy~\cite{Borsanyi:2020mff}. This remarkable result weakens the long-standing discrepancy between the muon $g$-2 SM prediction and the experimentally measured value. However, this result shows a tension with the time-like data-driven determinations of $a_{\mu}^{\rm HVP}({\rm LO})$, being $2.2 \sigma$ higher than the Muon $g$-2 Theory Initiative data-driven value. Moreover, shifts up of the $e^+ e^- \to $ hadrons cross section, due to unforeseen missing contributions, to increase $a_{\mu}^{\rm HVP}({\rm LO})$ and solve the present muon $g$-2 discrepancy, lead to conflicts with the global electroweak fit if they occur at energies higher than $\sim$1~GeV (and below that energy they are deemed improbable given the large required increases in the hadronic cross section)~\cite{Passera:2008jk,Keshavarzi:2020bfy,Crivellin:2020zul,Malaescu:2020zuc,Colangelo:2020lcg}. A new and competitive determination of $a_{\mu}^{\rm HVP}$, possibly at NNLO accuracy, based on a method alternative to the time-like and lattice QCD ones, is therefore desirable.

A novel approach to determine the leading hadronic contribution to the muon $g$-2, measuring the effective electromagnetic coupling in the space-like region via scattering data, was proposed a few years ago~\cite{CarloniCalame:2015obs}. The elastic scattering of high-energy muons on atomic electrons has been identified as an ideal process for this measurement, and a new experiment, MUonE, has been proposed at CERN to measure the shape of the differential cross section of muon-electron elastic scattering as a function of the space-like squared momentum transfer~\cite{Abbiendi:2016xup,MUonE:LoI,Banerjee:2020tdt}.

In this paper we investigate the HVP contributions to the muon $g$-2 in the space-like region. At LO, simple results are long known and form the basis for present lattice QCD and future MUonE determinations of $a_{\mu}^{\rm HVP}({\rm LO})$. Our goal is to provide simple analytic expressions to extend the space-like calculation of the $a_{\mu}^{\rm HVP}$ contribution to NNLO.

%%%%%%%%%%%%%%%%%%%%%%%%%%%%%%%%%%%%%%%%%%%%%%%%%%%%%%%%%%%%
\section{The HVP contribution at leading order}
\label{HVPLO}

%%%%%%%%%%%%%%%%%%%%%%%%%%
\subsection{Time-like method} 

Consider the hadronic component of the vacuum polarization (VP) tensor with four-momentum $q$,
\begin{align}
  i \Pi_{\rm h}^{\mu\nu}(q) &= i \Pi_{\rm h}(q^2) \left(g^{\mu\nu}q^2-q^\mu q^\nu \right)  \notag \\
                             &=  \int {\rm d}^4 x \,\, e^{iqx} \langle 0 | T \left\{j^\mu_{\rm em}(x) j^\nu_{\rm em} (0) \right\} | 0 \rangle,
\end{align}
where $j^\mu_{\rm em}(x)$ is the electromagnetic hadronic current and $\Pi_{\rm h}(q^2)$ is the renormalized HVP function satisfying the condition $\Pi_{\rm h}(0) = 0$. The function $\Pi_{\rm h}(q^2)$ cannot be calculated in perturbation theory because of the non-perturbative nature of the strong interactions at low energy. Yet, the optical theorem 
\begin{equation}
{\rm Im} \Pi_{\rm h}(s) = (\alpha/3) R(s),
\end{equation} 
where $\alpha$ is the fine-structure constant and the $R$-ratio is
\begin{equation}
  R(s) \,=\, \frac{\sigma(e^+e^- \to \text{hadrons})}{4\pi \alpha^2/(3 s)},
\end{equation}
allows to express the imaginary part of the hadronic vacuum polarization in terms of the measured cross section of the process $e^+ e^- \to $ hadrons as a function of the positive squared four-momentum transfer $s$. This result forms the basis for the time-like method.

\begin{figure}
\begin{center}
\includegraphics[width=0.8\columnwidth]{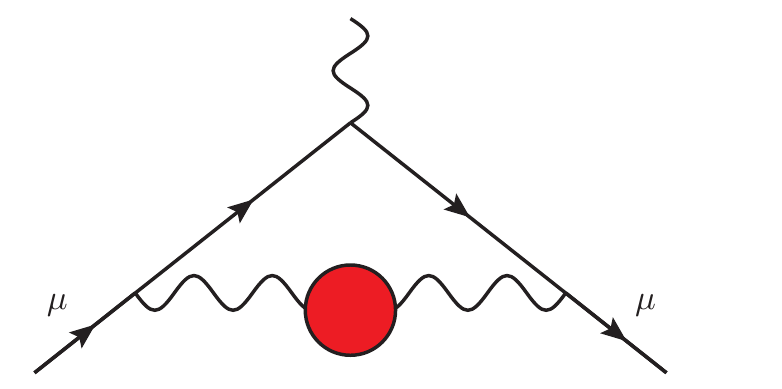}
\caption{The leading, $\mathcal{O}(\alpha^2)$, hadronic contribution to the muon $g$-2. The red blob indicates the HVP insertion.}
\label{fig:LOdiagram}
\end{center}
\end{figure}

The LO hadronic contribution to the muon $g$-2, due to the $\mathcal{O}(\alpha^2)$ diagram shown in Fig.~\ref{fig:LOdiagram}, can be calculated integrating experimental time-like (i.e.\ $q^2>0$) data using the well-known formula~\cite{Bouchiat:1961lbg,Brodsky:1967sr,Lautrup:1968tdb}
\begin{equation}
	a_{\mu}^{\rm HVP}({\rm LO}) = \frac{\alpha}{\pi^2} \int_{s_0}^{\infty} \frac{{\rm d}s}{s} 
	\;K^{(2)}(s/m^2)\; {\rm Im} \Pi_{\rm h} (s),
\label{eq:amuHLOtime}
\end{equation}
where $m$ is the muon mass and $s_0=m^2_{\pi^0}$ is the squared neutral pion mass. Defining
\begin{equation}
	z = \frac{q^2}{m^2}
	\label{eq:z}
\end{equation}
and the rationalizing variable
\begin{equation}	
         y(z) = \frac{z-\sqrt{z(z-4)}}{z+\sqrt{z(z-4)}},
\label{eq:y}
\end{equation}
the second-order function $K^{(2)}(z)$ for $z \geq 0$ is 
\begin{align}	
	K^{(2)}(z) &= \frac{1}{2} -z + \left( \frac{z^2}{2} -z \right) \ln z \nonumber \\
	&+ \frac{\ln y(z)}{\sqrt{z(z-4)}} \left( z -2 z^2 + \frac{z^3}{2}\right).
\label{eq:K2}
\end{align}
For $z \geq 0$, $K^{(2)}(z)$ is real, positive and monotonic (it has no cut for $0\leq z \leq 4$). At $z=0$, $K^{(2)}(0)=1/2$, while for $z \to +\infty$ the asymptotic behaviour of this kernel function is $K^{(2)}(z) \to 1/(3z)$, therefore vanishing at infinity. 

%%%%%%%%%%%%%%%%%%%%%%%%%%
\subsection{Space-like method}

The time-like expression for $a_{\mu}^{\rm HVP}({\rm LO})$ provided by~\eq{amuHLOtime} can be rewritten using the dispersion relation satisfied by $K^{(2)}(z)$~\cite{Barbieri:1974nc},
\begin{equation}	
	K^{(2)} (z) = \frac{1}{\pi} \int_{-\infty}^0 \! {\rm d}z'\, \frac{{\rm Im} K^{(2)} (z')}{z'-z}, \quad z>0.
\label{eq:DRK2t}
\end{equation}
Indeed, replacing $K^{(2)}(s/m^2)$ in~\eq{amuHLOtime} with~\eq{DRK2t} and integrating over $s$ via the subtracted dispersion relation satisfied by $\Pi_{\rm h}(q^2)$,
\begin{equation}
	\frac{\Pi_{\rm h}(q^2)}{q^2} =  \frac{1}{\pi}  \int_{s_0}^{\infty}  \frac{{\rm d}s}{s} \frac{{\rm Im} \Pi_{\rm h}(s)}{s-q^2}, \quad q^2<0,
\label{eq:DRPi}
\end{equation}
we obtain the space-like expression
\begin{equation}
	a_{\mu}^{\rm HVP}({\rm LO}) = - \frac{\alpha}{\pi^2} \int_{-\infty}^{0} \frac{{\rm d}t}{t} \,  \Pi_{\rm h}(t) \, {\rm Im} K^{(2)}(t/m^2).
\label{eq:amuHLOspace}
\end{equation}

The function $K^{(2)}(z)$, real for any $z \geq 0$, has a cut along the negative real axis $z<0$ with the imaginary part
\begin{align}	
{\rm Im} K^{(2)} (z + i \epsilon) &= \pi \, \theta(-z) \! \left[  \frac{z^2}{2} -z  + \frac{z -2 z^2 + z^3/2}{\sqrt{z(z-4)}}\right] \nonumber \\
						&= \pi \, \theta(-z) \, F^{(2)}(1/y(z)),
\label{eq:ImK2}
\end{align}	
where 
\begin{equation}	
	F^{(2)} (u)  = \frac{u+1}{u-1} \, u^2.
\label{eq:F2}
\end{equation}
The $i \epsilon$ prescription, with $\epsilon>0$, indicates that, in correspondence of the cut, the function Im$K^{(2)}(z)$ is evaluated approaching the real axis from above.

If in~\eq{amuHLOspace} one uses the explicit expression for ${\rm Im} K^{(2)}\!(t/m^2)$ of~\eq{ImK2} and changes the integration variable from $t$ to $x=1+1/y$ via the substitution 
\begin{equation}
	t(x) = \frac{m^2 x^2}{x-1},
\label{eq:tofx}
\end{equation}
obtained from \eq{y}, one finds~\cite{Lautrup:1971jf}
\begin{equation}
	a_{\mu}^{\rm HVP}({\rm LO}) = \frac{\alpha}{\pi} \int_{0}^{1} {\rm d}x \, \kappa^{(2)} (x) \, \Delta \alpha_{\rm h}( t(x) ),
\label{eq:amuHLOspacex}
\end{equation}
where the space-like kernel is remarkably simple,
\begin{equation}
	\kappa^{(2)} (x) = 1-x
\label{eq:kappa2}
\end{equation}
and $\Delta \alpha_{\rm h} (t)= -\Pi_{\rm h}(t)$ is the (five-flavor) hadronic contribution to the running of the electromagnetic coupling in the space-like region, $\alpha (t) = \alpha/(1- \Delta \alpha (t))$.

Equation~(\ref{eq:amuHLOspacex}) (or forms equivalent to it) is used in lattice QCD calculations of $a_{\mu}^{\rm HVP}({\rm LO})$ (see e.g.\ \cite{Blum:2002ii} and a discussion in~\cite{Aoyama:2020ynm}) and forms the basis for the MUonE proposal to determine $a_{\mu}^{\rm HVP}({\rm LO})$ via muon-electron scattering data~\cite{CarloniCalame:2015obs,Abbiendi:2016xup,MUonE:LoI,Banerjee:2020tdt}. 

We close this Section noting that, in Fig.~\ref{fig:LOdiagram}, a virtual photon can be emitted and reabsorbed by the HVP insertion of the LO diagram. These irreducible hadronic contributions, although of higher order in $\alpha$, are normally incorporated into the time-like determination of $a_{\mu}^{\rm HVP}({\rm LO})$ via the inclusion of final-state radiation corrections in the $R$-ratio (see e.g.~\cite{Aoyama:2020ynm,Jegerlehner:2017gek}).\footnote{Note that, consistently, the lower limit of integration in~\eq{amuHLOtime} has been chosen to be $s_0=m^2_{\pi^0}$, the threshold of the $\pi^0 \gamma$ cross section.} For a comparison, also space-like evaluations of $a_{\mu}^{\rm HVP}({\rm LO})$ should therefore incorporate these higher-order corrections, including them in $\Delta \alpha_{\rm h}(t)$ in~\eq{amuHLOspacex}. In this respect, the fully inclusive measurement of $\Delta \alpha_{\rm h}(t)$ expected from MUonE is ideal~\cite{Fael:2019nsf}.

%%%%%%%%%%%%%%%%%%%%%%%%%%%%%%%%%%%%%%%%%%%%%%%%%%%%%%%%%%%%
\section{The HVP contribution at NLO}
\label{HVPNLO}

\begin{figure}
\begin{center}
\includegraphics[width=\columnwidth]{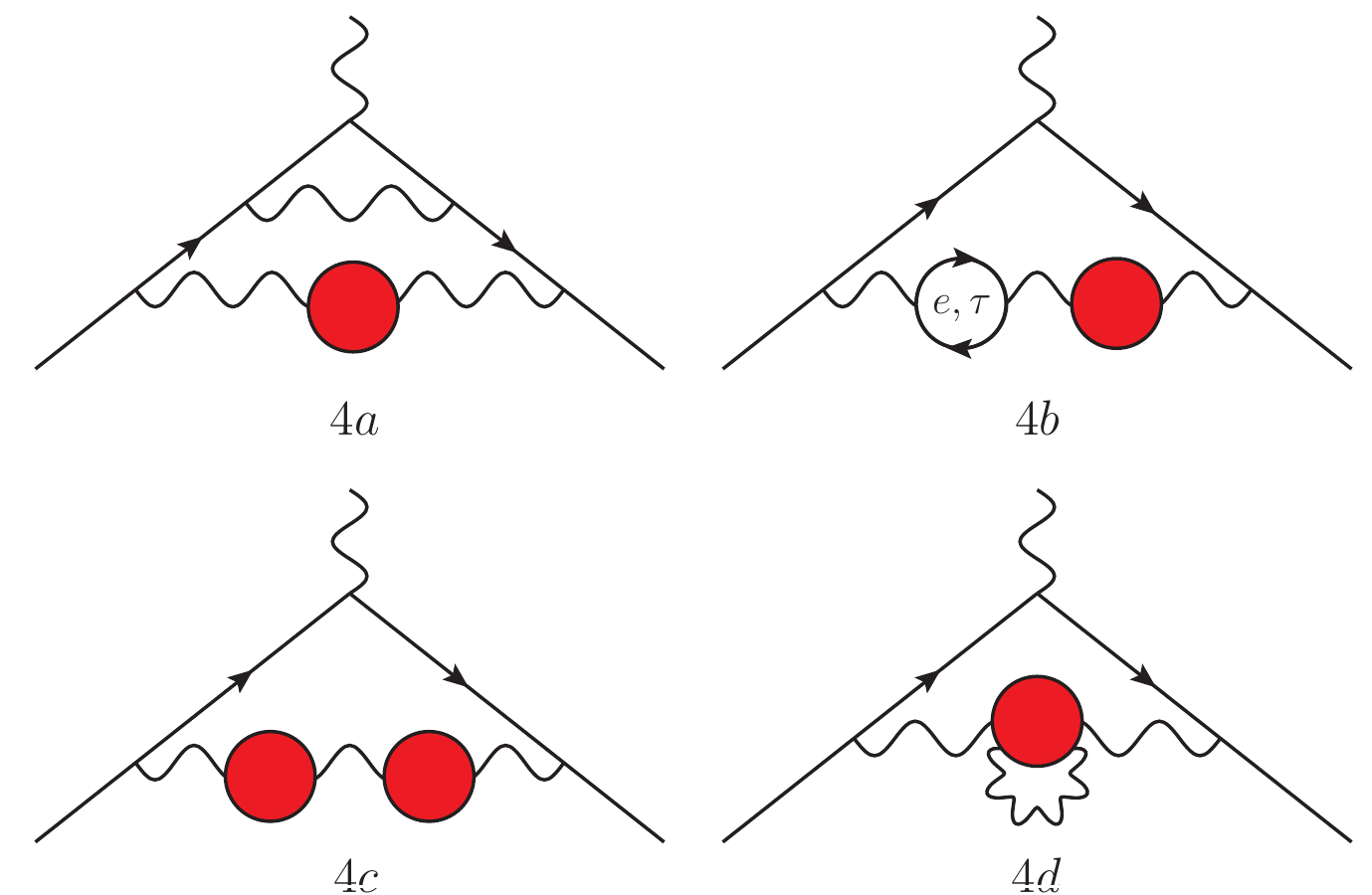}
\caption{Sample $\mathcal{O}(\alpha^3)$ diagrams contributing to the HVP corrections to the muon $g$-2.}
\label{fig:NLOdiagrams}
\end{center}
\end{figure}

The hadronic vacuum polarization contribution to the muon $g$-2 at NLO, $a_{\mu}^{\rm HVP}({\rm NLO})$ has been studied as early as in Ref.~\cite{Calmet:1976kd}. It is due to $\mathcal{O}(\alpha^3)$ diagrams that can be classified as follows (see Fig.~\ref{fig:NLOdiagrams}).
Class $(4a)$ comprises diagrams with one single HVP insertion in one of the photon lines of the two-loop QED diagrams contributing to the muon $g$-2, without any VP insertion due to electron or tau loops. Class $(4b)$ contains diagrams with one HVP and one additional VP due to an electron or tau loop. Class $(4c)$ consists of the single diagram with two HVPs. Class $(4d)$ diagrams contain internal radiative corrections to the HVP. As discussed in the previous Section, this contribution is not considered as part of $a_{\mu}^{\rm HVP}({\rm NLO})$, although of the same order in $\alpha$, because it is already incorporated into $a_{\mu}^{\rm HVP}({\rm LO})$. 
Analogously, the $\mathcal{O}(\alpha^4)$ contributions obtained by adding to the diagrams of classes $(4a)$, $(4b)$ and $(4c)$ a virtual photon emitted and reabsorbed by an HVP insertion, although of higher order in $\alpha$, should be incorporated into $a_{\mu}^{\rm HVP}({\rm NLO})$, either via the $R$-ratio (in the time-like approach) or via $\Delta \alpha_{\rm h}(t)$ (in the space-like one). If a second virtual photon is attached to the HVP insertion of class $(4d)$, the resulting contribution should be incorporated into $a_{\mu}^{\rm HVP}({\rm LO})$ (see also Section~\ref{HVPNNLO}).

Numerically, class $(4a)$ yields the largest (negative) contribution, class $(4b)$ partially cancels it, and class $(4c)$ is small, as expected. Their sum
\begin{equation}
	a_{\mu}^{\rm HVP}({\rm NLO}) = a_{\mu}^{(4a)} + a_{\mu}^{(4b)} + a_{\mu}^{(4c)}
\label{eq:amuHVPNLO}
\end{equation}
is negative and of $\mathcal{O}(10^{-9})$.

%%%%%%%%%%%%%%%%%%%%%%%%%%
\subsection{Class $(4a)$}
\label{HVPNLOa}

The NLO HVP contribution of class $(4a)$ to the muon $g$-2 can be written in the time-like form~\cite{Barbieri:1974nc}
\begin{equation}
	a_{\mu}^{(4a)} =  \frac{\alpha^2}{\pi^3} \int_{s_0}^{\infty} \frac{{\rm d}s}{s} \, 2K^{(4)}(s/m^2)\, {\rm Im} \Pi_{\rm h} (s).
\label{eq:amuAtime}
\end{equation}
The fourth-order function $K^{(4)}(z)$ was first computed by Barbieri and Remiddi in~\cite{Barbieri:1974nc}.\footnote{Note the coefficient 2 in front of the function $K^{(4)}(z)$ due to the original normalization chosen in Ref.~\cite{Barbieri:1974nc}.} Its lengthy expression is reported in their Eq.~(3.21) for $z>0$, where it is real and negative. An approximate series expansion for $K^{(4)}(s/m^2)$ in the parameter $m^2/s$, with terms up to fourth order, can be found in~\cite{Krause:1996rf}.

Like $K^{(2)}(z)$, the function $K^{(4)}(z)$ is real for any $z \geq 0$, has a cut for $z<0$, and satisfies the dispersion relation
\begin{equation}	
	K^{(4)} (z) = \frac{1}{\pi} \int_{-\infty}^0 \! {\rm d}z'\, \frac{{\rm Im} K^{(4)} (z')}{z'-z}, \quad z>0.
\label{eq:DRK4t}
\end{equation}
Just as we did for $a_{\mu}^{\rm HVP}({\rm LO})$, using the dispersion relations~(\ref{eq:DRPi}) and (\ref{eq:DRK4t}) the NLO hadronic contribution of class $(4a)$ can be cast in the space-like form
\begin{equation}
	a_{\mu}^{(4a)} = - \frac{\alpha^2}{\pi^3} \int_{-\infty}^{0} \frac{{\rm d}t}{t} \,  \Pi_{\rm h}(t) \, 2{\rm Im} K^{(4)}(t/m^2).
\label{eq:amuAspace}
\end{equation}

The function ${\rm Im} K^{(4)}(t/m^2)$ can be calculated from the $K^{(4)}(z)$ expression of Ref.~\cite{Barbieri:1974nc}. Taking the imaginary parts of the polylogarithms of order 1, 2, and 3, we obtain\footnote{After presenting our ${\rm Im}  K^{(4)} (z)$ result, Eqs.(\ref{eq:ImK4}--\ref{eq:R15}), in~\cite{Laporta:Strong2021} (see also~\cite{Passera:Remiddi2021}), we were informed by Alexander Nesterenko that he has independently derived it in~\cite{Nesterenko:2021byp}.}  
\begin{equation}	
	{\rm Im}  K^{(4)} (z + i \epsilon) = \pi \, \theta(-z) \, F^{(4)} (1/y(z)),
\label{eq:ImK4}
\end{equation}
where
\begin{align}	
	F^{(4)} (u)    & = R_1(u) + R_2(u) \ln (-u) \nonumber \\
	          & \,+ R_3(u) \ln (1+u) + R_4(u) \ln (1-u) \nonumber \\
	          & \,+ R_5(u)\big[ 4{\rm Li}_2(u) + 2{\rm Li}_2(-u)   \nonumber \\ 
	&~~~~+\ln(-u) \ln \! \left( (1-u)^2  (1+u)  \right) \! \big],
\label{eq:F}
\end{align}
and the rational functions $R_i(u)$ $(i=1,\ldots,5)$ are 
\begin{align}
	R_1 \!= &  \frac{23 u^6 \!-\! 37 u^5 \!+\! 124 u^4 \!-\! 86 u^3 \!-\! 57 u^2 \!+\! 99 u \!+\! 78}{72 (u-1)^2 u (u+1)}, \nonumber \\
	R_2 \!= &  \frac{12 u^8 \!-\! 11 u^7 \!-\! 78 u^6 \!+\! 21 u^5 \!+\! 4 u^4 \!-\! 15 u^3 \!+\! 13 u \!+\! 6}{12 (u-1)^3 u (u+1)^2},  \nonumber \\
	R_3 \!= &  \frac{(u+1) \left(-u^3 + 7 u^2 + 8 u + 6\right)}{12 u^2}, \nonumber \\
	R_4 \!= &  \frac{-7 u^4 - 8 u^3 + 8 u + 7}{12 u^2}, \nonumber \\
	R_5 \!= & -\frac{3 u^4 + 5 u^3 + 7 u^2 + 5 u + 3}{6 u^2}.
\label{eq:R15}
\end{align}
The dilogarithm is ${\rm Li}_2(u) = -\int_{0}^u ({\rm d}v/v) \ln (1-v)$.

Using the explicit expression for ${\rm Im} K^{(4)}\!(t/m^2)$ of~\eq{ImK4},~\eq{amuAspace} can be conveniently expressed in terms of the variable $x=1+1/y$. We obtain
\begin{equation}
	a_{\mu}^{(4a)}  =  \left( \frac{\alpha}{\pi} \right)^2 \! \int_{0}^{1} {\rm d}x \, \kappa^{(4)} (x) \, \Delta \alpha_{\rm h}( t(x) ),
\label{eq:amuAspacex}
\end{equation}
where 
\begin{equation}
	\kappa^{(4)} (x) = \frac{2-x}{x\left(x-1\right)} \, 2F^{(4)} (x-1).
\label{eq:kappa4}
\end{equation}
For $0 \leq x < 1$, $z \leq 0$. Equation~(\ref{eq:amuAspacex}) is the analogue of~\eq{amuHLOspacex} for the NLO contribution of class $(4a)$.

\begin{figure}
\begin{center}
\includegraphics[width=\columnwidth]{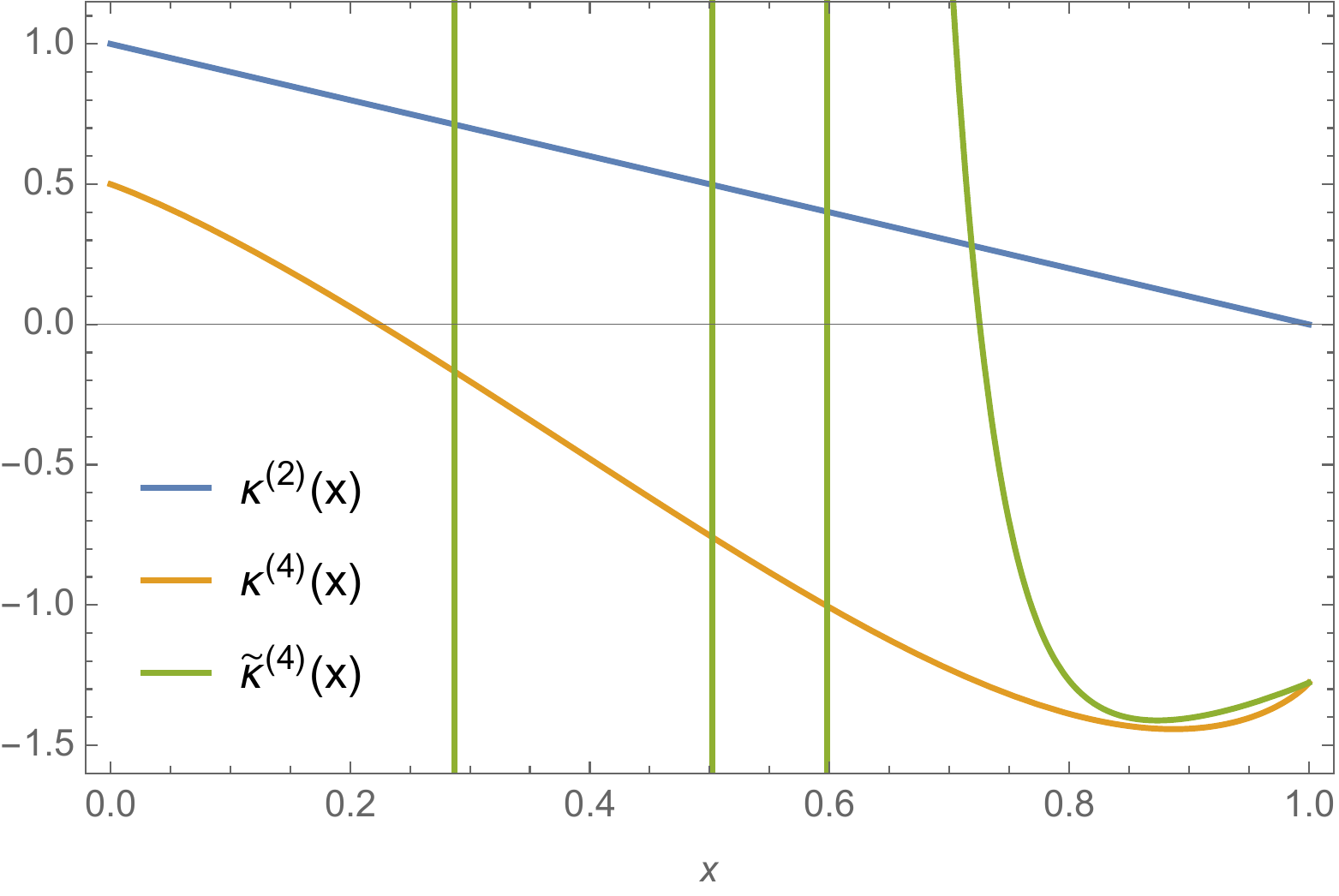}
\caption{The space-like functions $\kappa^{(2)} (x)$ (blue), $\kappa^{(4)} (x)$ (orange) and $\tilde{\kappa}^{(4)} (x)$ (green).}
\label{fig:kappa24tilde}
\end{center}
\end{figure}

\begin{figure}
\begin{center}
\includegraphics[width=\columnwidth]{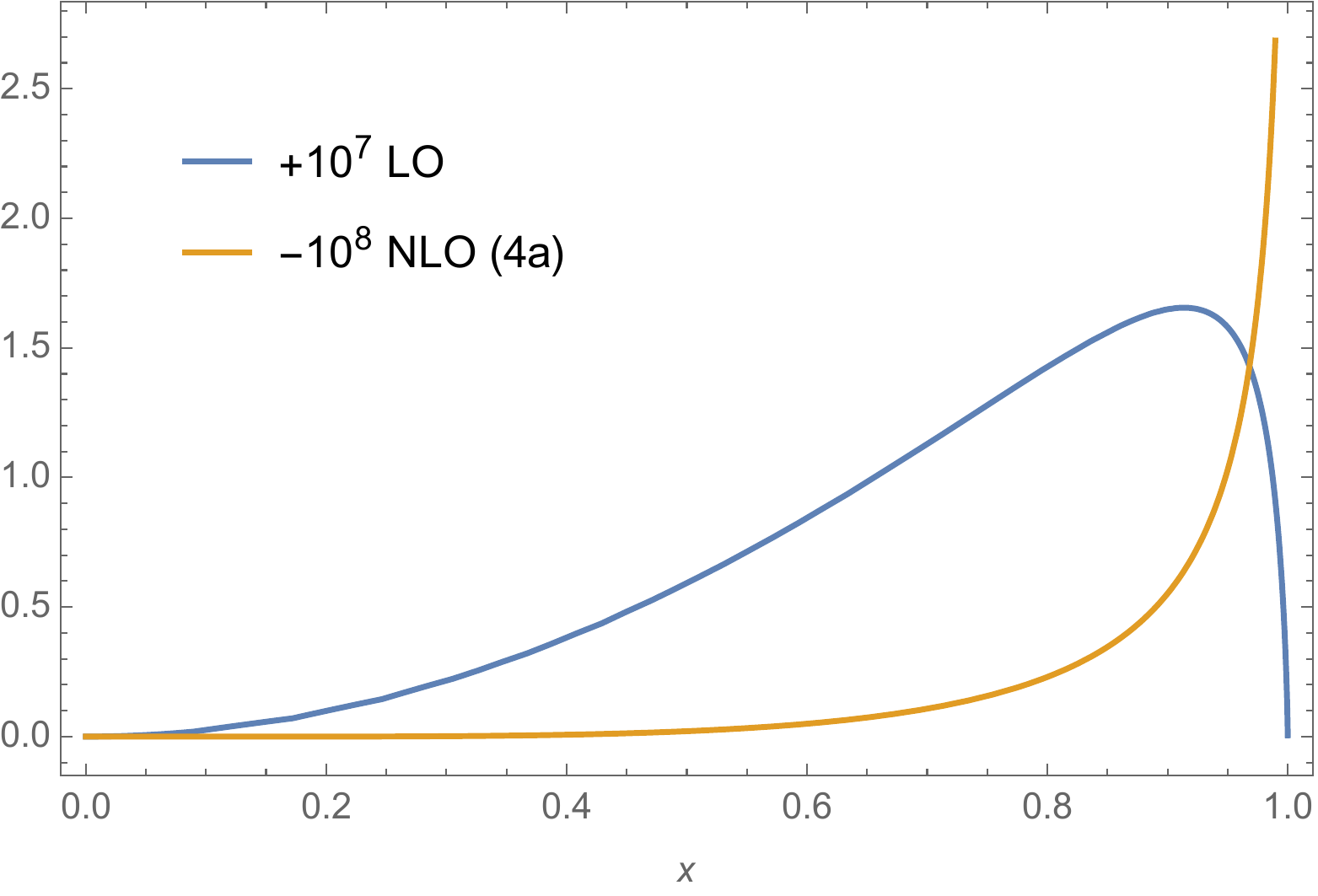}
\caption{The integrands $(\alpha/\pi) \kappa^{(2)} (x) \Delta \alpha_{\rm h}(t(x))$ (blue) and $(\alpha/\pi)^2 \kappa^{(4)} (x) \Delta \alpha_{\rm h}(t(x))$ (orange) of Eqs.(\ref{eq:amuHLOspacex}) and (\ref{eq:amuAspacex}), multiplied by $10^7$ and $-10^8$, respectively.}
\label{fig:integrand24}
\end{center}
\end{figure}

Figure~\ref{fig:kappa24tilde} shows the space-like functions $\kappa^{(2)} (x)$ and $\kappa^{(4)} (x)$ entering the $a_{\mu}^{\rm HVP}({\rm LO})$ and $a_{\mu}^{(4a)}$ expressions, respectively. We note that the function $\kappa^{(4)} (x)$ provides a stronger weight to $\Delta \alpha_{\rm h}(q^2)$ at large negative values of $q^2$ than $\kappa^{(2)} (x)$. In particular, for $q^2 \to -\infty$, $\kappa^{(2)} (1) =0$, whereas $\kappa^{(4)} (1) = -23/18$. 
Figure~\ref{fig:integrand24} shows the LO integrand $(\alpha/\pi) \kappa^{(2)} (x) \Delta \alpha_{\rm h}(t(x))$ of~\eq{amuHLOspacex} and the NLO integrand $(\alpha/\pi)^2 \kappa^{(4)} (x) \Delta \alpha_{\rm h}(t(x))$ of~\eq{amuAspacex}, multiplied by $10^7$ and $-10^8$, respectively. The LO integrand has a peak at $x \sim 0.914$, where $t \sim -(0.33 {\rm GeV})^2$. On the other hand, the NLO integrand of class $(4a)$ increases monotonically with $x \to 1$ (i.e.\ with $t \to -\infty$) like $\sim \ln (1-x)$.

An approximate expression for the space-like formula in~\eq{amuAspacex} was provided in Ref.~\cite{Chakraborty:2018iyb}. To obtain it, the authors started considering the approximate fourth-order series expansion of Ref.~\cite{Krause:1996rf} for the time-like function $K^{(4)}(s/m^2)$ in the small parameter $r=m^2/s$. This series expansion contains only powers $r^n$ of degree $n=1,2,3,4$, multiplied by constants, $\ln r$, and $(\ln r)^2$ terms. Then, as suggested in~\cite{Groote:2001vu}, they exploited generating integral representations to fit the $r^n$ and $r^n \! \ln r$ terms of the approximate fourth-order series expansion for $K^{(4)}(1/r)$, but not the $r^n (\ln r)^2$ ones, and used the usual dispersion relation satisfied by $\Pi_{\rm h}(q^2)$ to perform the integral over $s$. After simple changes of variables, their approximation can be compared with our exact function $\kappa^{(4)} (x)$. We repeated the analysis of Ref.~\cite{Chakraborty:2018iyb} confirming their approximate result (in particular, their Eqs.~(A1,A2)) which, translated in our notation, is called here $\tilde{\kappa}^{(4)} (x)$.

The approximate function $\tilde{\kappa}^{(4)} (x)$ is plotted in Fig.~\ref{fig:kappa24tilde} (indicated by the green line). While the exact function $\kappa^{(4)} (x)$ varies smoothly over the entire region $0 \leq x \leq 1$, $\tilde{\kappa}^{(4)} (x)$ strongly oscillates, leading to large numerical cancelations when employed in the integral of~\eq{amuAspacex} instead of $\kappa^{(4)} (x)$. Using the Fortran libraries KNT18VP~\cite{Harlander:2002ur,Hagiwara:2003da,Hagiwara:2006jt,WorkingGrouponRadiativeCorrections:2010bjp,Hagiwara:2011af,Keshavarzi:2018mgv} for the numerical implementation of $\Delta \alpha_{\rm h}(t(x))$ in the space-like region, we computed two numerical values for $a_{\mu}^{(4a)}$ in~\eq{amuAspacex}: one obtained using the exact function $\kappa^{(4)} (x)$ and a second one obtained replacing $\kappa^{(4)} (x)$ with the approximate $\tilde{\kappa}^{(4)} (x)$. The two values differ by about 3\%. Adding to $a_{\mu}^{(4a)}$ the contributions $a_{\mu}^{(4b)}$ and $a_{\mu}^{(4c)}$ (discussed later), the total $a_{\mu}^{\rm HVP}({\rm NLO})$ contribution computed using the $\tilde{\kappa}^{(4)} (x)$ approximation differs from the one computed via our exact function $\kappa^{(4)} (x)$ by about 6\%.

It is interesting to investigate the source of the above $\sim 6$\% discrepancy. To improve the $\tilde{\kappa}^{(4)} (x)$ approximation, we proceeded in two directions. The first one consisted in repeating the analysis of Ref.~\cite{Chakraborty:2018iyb}, starting however from higher-order series expansions for the exact $K^{(4)}(s/m^2)$ function of Barbieri and Remiddi~\cite{Barbieri:1974nc} (we considered $n$ up to $n_{\rm max}=24$), rather than from the fourth-order (i.e.\ $n_{\rm max}=4$) series expansion for $K^{(4)}(s/m^2)$ of Ref.~\cite{Krause:1996rf}. Our second improvement consisted in exploiting generating integral representations to fit the $r^n$, $r^n \! \ln r$, as well as the $r^n (\ln r)^2$ terms which were omitted in the analysis of Ref.~\cite{Chakraborty:2018iyb}. Our studies show that the inclusion of the $r^n (\ln r)^2$ terms greatly improves the $\kappa^{(4)} (x)$ approximations, even if the order $n_{\rm max}$ of the series expansion for $K^{(4)}(s/m^2)$ is not increased above four. Calling $\bar{\kappa}^{(4)} (x,n_{\rm max})$ our improved approximations to $\kappa^{(4)} (x)$, obtained including $r^n (\ln r)^2$ terms and starting from series expansions for $K^{(4)}(s/m^2)$ up to order $n_{\rm max}$, we verified that the total $a_{\mu}^{\rm HVP}({\rm NLO})$ contribution computed using our $\bar{\kappa}^{(4)} (x,4)$ differs by less than one per mille from the one computed via our exact function $\kappa^{(4)} (x)$. Even better agreements were reached increasing the order $n_{\rm max}$.

The authors of Ref.~\cite{Chakraborty:2018iyb} added an $O(10\%)$ uncertainty to their final result to take into account the error induced by the omission of the $r^n (\ln r)^2$ terms. This uncertainty, which dominates the error of their final result, can be eliminated using the exact formula for $\kappa^{(4)} (x)$ provided in this paper.

%%%%%%%%%%%%%%%%%%%%%%%%%%
\subsection{Classes $(4b)$ and $(4c)$}
\label{HVPNLObc}

The space-like expressions for the contributions of classes $(4b)$ and $(4c)$ to the muon $g$-2 are~\cite{Jegerlehner:2017gek,Chakraborty:2018iyb}

\begin{align}
	a_{\mu}^{(4b)} &=  \frac{\alpha}{\pi} \int_{0}^{1} {\rm d}x \, \kappa^{(2)} (x) \, \Delta \alpha_{\rm h}( t(x) ) 
	\label{eq:amuBspacex} \nonumber \\ 
	                      & ~~~~~~~ \times 2 \left[ \Delta \alpha_e^{(2)}( t(x) )  + \Delta \alpha_{\rm \tau}^{(2)}( t(x) ) \right], \\
	a_{\mu}^{(4c)} & =  \frac{\alpha}{\pi} \int_{0}^{1} {\rm d}x \, \kappa^{(2)} (x) \, \left[ \Delta \alpha_{\rm h}( t(x) ) \right]^2,
\label{eq:amuCspacex}
\end{align}
where $\Pi_{\ell}^{(2)} (t) = -\Delta \alpha_{\ell}^{(2)}(t)$ is the renormalized one-loop QED VP function in the space-like region, with a lepton $\ell=e,\tau$ of mass $m_\ell$ in the loop,
\begin{equation}
	\Pi_{\ell}^{(2)} (t) = \frac{\alpha}{\pi}  \left[ 
					\frac{8}{9}  - \frac{\beta_{\ell}^2}{3}  + \beta_{\ell} \! \left(\frac{1}{2} - \frac{\beta^2_{\ell}}{6} \right) \! \ln \frac{\beta_{\ell}-1}{\beta_{\ell}
					 +1} \right]
\end{equation}
and $\beta_{\ell}= \sqrt{1-4m_\ell^2/t}$. Equations (\ref{eq:amuBspacex},\ref{eq:amuCspacex}) can be immediately obtained from the time-like formulae of Ref.~\cite{Krause:1996rf} using the usual dispersion relation satisfied by $\Pi_{\rm h}(t)$ and $\Pi_{\ell}^{(2)} (t)$ to perform the integrals over $s$~\cite{Lautrup:1971jf,Chakraborty:2018iyb}.

\begin{figure}
\begin{center}
\includegraphics[width=\columnwidth]{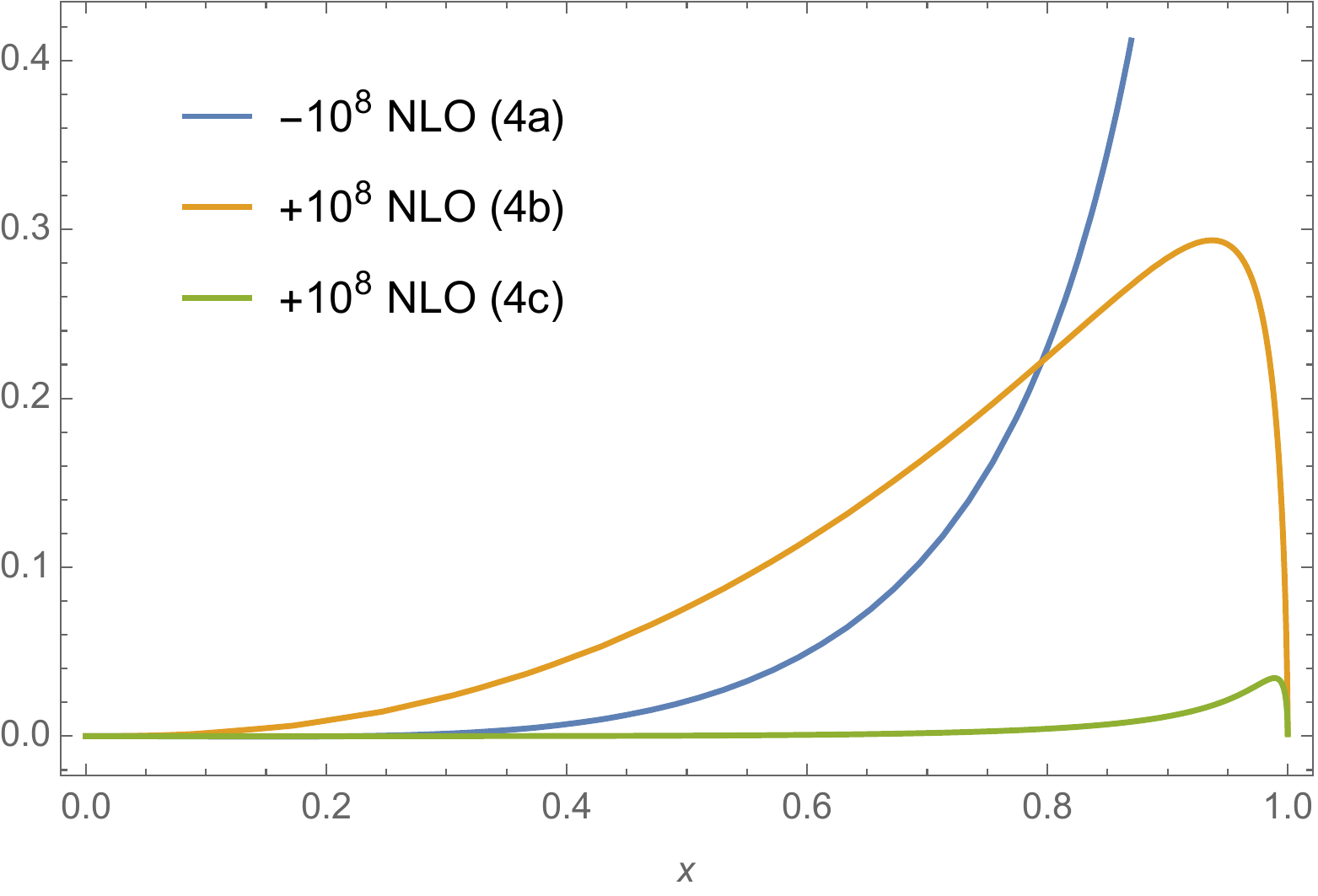}
\caption{The NLO integrands of~Eqs.~(\ref{eq:amuAspacex}) (blue), (\ref{eq:amuBspacex}) (orange), and (\ref{eq:amuCspacex}) (green), multiplied by 
$-10^8$,
$10^8$, and
$10^8$, respectively.}
\label{fig:NLOabc}
\end{center}
\end{figure}

Figure~\ref{fig:NLOabc} shows the NLO integrands of~Eqs.~(\ref{eq:amuAspacex}), (\ref{eq:amuBspacex}), and (\ref{eq:amuCspacex}), multiplied by 
$-10^8$,
$10^8$, and
$10^8$, respectively. 

%%%%%%%%%%%%%%%%%%%%%%%%%%%%%%%%%%%%%%%%%%%%%%%%%%%%%%%%%%%%
\section{The HVP contribution at NNLO}
\label{HVPNNLO}
 
\begin{figure*}
\begin{center}
\includegraphics[width=\textwidth]{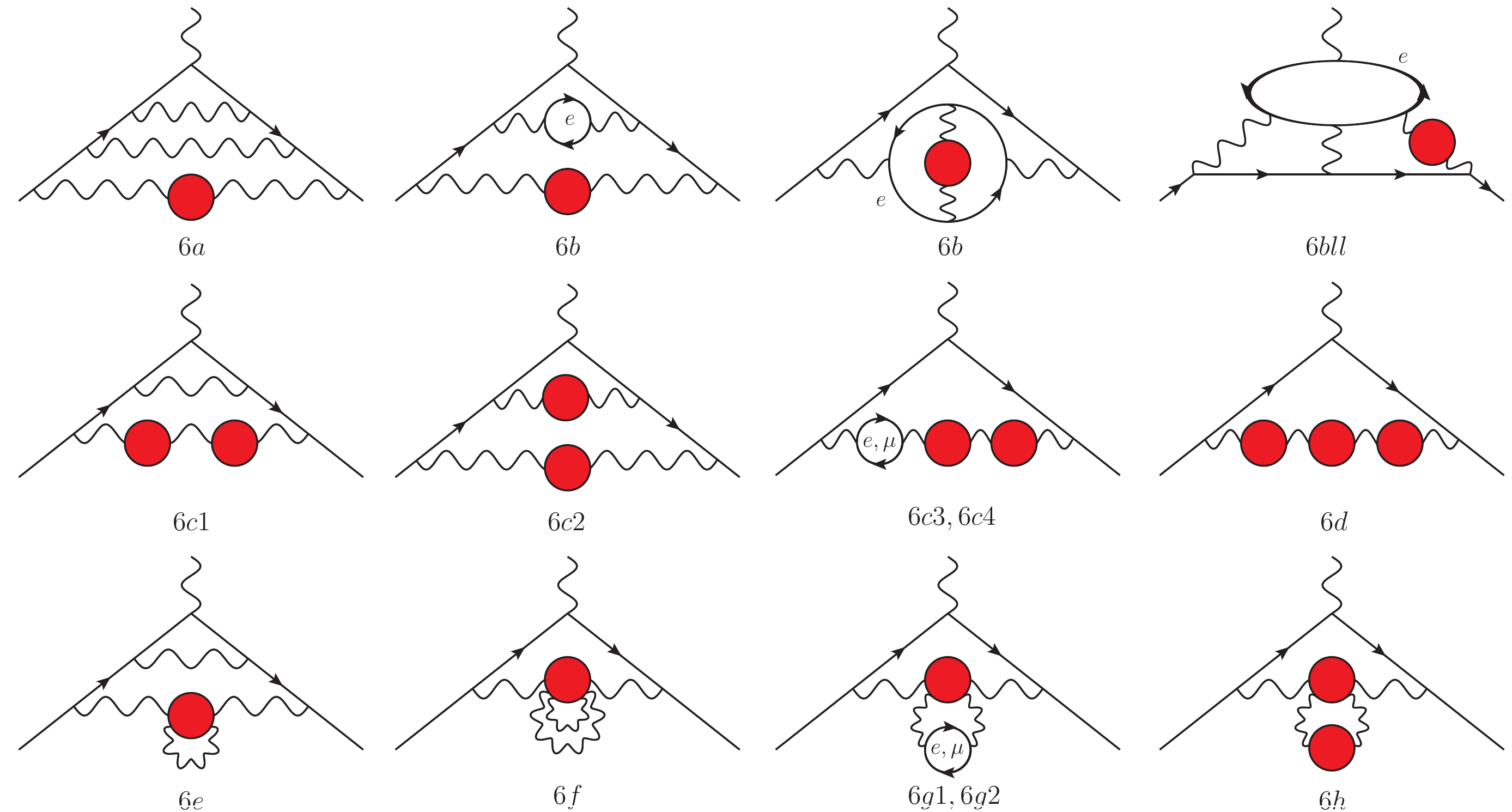}
\caption{Sample $\mathcal{O}(\alpha^4)$ diagrams contributing to the HVP corrections to the muon $g$-2.}
\label{fig:NNLOdiagrams}
\end{center}
\end{figure*}
The hadronic vacuum polarization contribution to the muon $g$-2 at NNLO, $a_{\mu}^{\rm HVP}({\rm NNLO})$, is due to several diagrams of $\mathcal{O}(\alpha^4)$. We divide them into the following classes (see Fig.~\ref{fig:NNLOdiagrams}).\footnote{At NNLO we neglect the contribution of tau loops as it is estimated to be smaller than $O(10^{-12})$~\cite{Kurz:2014wya}.} Class $(6a)$ contains diagrams with one HVP insertion and up to two photons added to the LO QED Feynman graph; it also includes diagrams with one or two muon VP loops and the light-by-light graph with a muon loop. Class $(6b)$ comprises diagrams with one HVP insertion and one or two electron VP loops and additional photonic or muon VP corrections; it also includes diagrams with one electron VP loop with an HVP insertion inside it. Class $(6bll)$ diagrams have one HVP insertion and light-by-light graphs with an electron loop; in these diagrams, the external photon couples to the electron. Class $(6c)$ contains diagrams with two HVP insertions and additional photonic corrections and/or electron or muon VP loops. Class $(6d)$ consists of the diagram with three HVP insertions. All of these classes were studied in Ref.~\cite{Kurz:2014wya} in the time-like approach.

Class $(6e)$ diagrams are obtained by adding to those of classes $(4a)$, $(4b)$ and $(4c)$ a virtual photon emitted and reabsorbed by an HVP insertion. As discussed in the previous Section, their contribution should not be considered as part of $a_{\mu}^{\rm HVP}({\rm NNLO})$, although of the same order in $\alpha$, because it is already incorporated into $a_{\mu}^{\rm HVP}({\rm NLO})$ via the $R$-ratio (in the time-like approach) or via $\Delta \alpha_{\rm h}(t)$ (in the space-like one). 
Similarly, the corrections of class $(6f)$, where two photons are emitted and reabsorbed by the HVP insertion of the LO diagram, should be already included in $a_{\mu}^{\rm HVP}({\rm LO})$. The impact of class $(6f)$ can be roughly estimated considering the corresponding class of diagrams where the HVP insertion is replaced by a muon VP; that four-loop QED contribution to the muon $g$-2 is $1.44 \times 10^{-12}$~\cite{Laporta:2017okg}. The effect of class $(6f)$ can thus be roughly estimated to be of $\mathcal{O}(10^{-12})$.
The contribution of class $(6g1)$ was recently studied in Ref.~\cite{Hoferichter:2021wyj}, where it was estimated to be $\lesssim 1 \times 10^{-11}$. 
The impact of classes $(6g2)$ and $(6h)$ can be roughly estimated, once again, considering the corresponding four-loop QED contribution to the muon $g$-2 where the HVP insertions are replaced by muon VPs: $3.24 \times 10^{-13}$~\cite{Laporta:2017okg}. The effect of classes $(6g2)$ and $(6h)$ can thus be roughly estimated to be of $\mathcal{O}(10^{-13})$. Classes $(6f)$, $(6g1)$, $(6g2)$ and $(6h)$ should be incorporated into $a_{\mu}^{\rm HVP}({\rm LO})$.

The sum of the NNLO contributions is, therefore,
\begin{align}
	a_{\mu}^{\rm HVP}({\rm NNLO}) = a_{\mu}^{(6a)} + a_{\mu}^{(6b)} + a_{\mu}^{(6bll)} + a_{\mu}^{(6c)} + a_{\mu}^{(6d)}.
\label{eq:amuHVPNLO}
\end{align}
It is positive and of $\mathcal{O}(10^{-10})$~\cite{Kurz:2014wya}.

%%%%%%%%%%%%%%%%%%%%%%%%%%
\subsection{Class $(6a)$}
\label{HVPNNLOa}

The contribution of class $(6a)$ can be written in the time-like form~\cite{Kurz:2014wya}
\begin{equation}
	a_{\mu}^{(6a)} =  \frac{\alpha^3}{\pi^4} \int_{s_0}^{\infty} \frac{{\rm d}s}{s} \, K^{(6a)}(s/m^2)\, {\rm Im} \Pi_{\rm h} (s).
\label{eq:amu6Atime}
\end{equation}
The sixth-order function $K^{(6a)}(z)$ is not known in exact form, but an approximate series expansion in the parameter $r=m^2/s$, with terms up to fourth order, was computed in~\cite{Kurz:2014wya}. This expansion contains powers $r^n$ of degree $n=1,2,3,4$, multiplied by constants, $\ln r$, $(\ln r)^2$ and $(\ln r)^3$ terms. Following a procedure similar to that described at NLO, we exploited generating integral representations to fit all the $r^n$, $r^n \! \ln r$, $r^n (\ln r)^2$, and $r^n (\ln r)^3$ terms of the $K^{(6a)}(s/m^2)$ expansion,
\begin{equation}
  K^{(6a)}(s/m^2)  =  r  \! \int_0^1 \! \! {\rm d} \xi  \!  \left[ \frac{L^{(6a)}(\xi)}{\xi +r} + \frac{P^{(6a)}(\xi)}{1+ r \xi} \right]
\label{eq:K6A}
\end{equation}
where
\begin{equation}
  L^{(6a)}(\xi)	=  G^{(6a)}(\xi) + H^{(6a)}(\xi) \ln \xi + J^{(6a)}(\xi) \ln^2 \! \xi
\label{eq:L6A}
\end{equation} 
and
\begin{align}
  G^{(6a)}(\xi)	&= g_0^{(6a)} \! + g_1^{(6a)} \xi + g_2^{(6a)} \xi^2 \! + g_3^{(6a)} \xi^3,\notag \\
  H^{(6a)}(\xi)	&= h_0^{(6a)} \! + h_1^{(6a)} \xi + h_2^{(6a)} \xi^2 \! + h_3^{(6a)} \xi^3, \notag \\
  J^{(6a)}(\xi)	&= j_0^{(6a)}  \! + j_1^{(6a)} \xi   + j_2^{(6a)} \xi^2 \! + j_3^{(6a)} \xi^3, \notag \\
  P^{(6a)}(\xi)	&= p_0^{(6a)} \! + p_1^{(6a)} \xi + p_2^{(6a)} \xi^2 \! + p_3^{(6a)} \xi^3,
  \end{align}
obtaining the coefficients $g_i^{(6a)}$, $h_i^{(6a)}$, $j_i^{(6a)}$ and $p_i^{(6a)}$ ($i=0,1,2,3$) reported in Table~\ref{table:6a}.

Inserting the integral representation of~\eq{K6A} in~\eq{amu6Atime}, the integral over $s$ can be performed using the dispersion relation satisfied by $\Pi_{\rm h}(q^2)$. With simple changes of variables we obtain
\begin{equation}
	a_{\mu}^{(6a)}  =  \left( \frac{\alpha}{\pi} \right)^3 \! \int_{0}^{1} {\rm d}x \, \bar{\kappa}^{(6a)} (x) \, \Delta \alpha_{\rm h}( t(x) ),
\label{eq:amu6Aspacex}
\end{equation}
where, for $0<x<x_{\mu}=(\sqrt 5 -1)/2=0.618\ldots$,
\begin{equation}
	\bar{\kappa}^{(6a)} (x) \,=\,  \frac{2-x}{x\left(1-x\right)} \, P^{(6a)} \!\!  \left(\frac{x^2}{1-x}\right),
\label{eq:kappa6alow}
\end{equation}
whereas, for $x_{\mu}<x<1$,
\begin{equation}
	\bar{\kappa}^{(6a)} (x) \,=\, \frac{2-x}{x^3} \, L^{(6a)} \!\!  \left(\frac{1-x}{x^2}\right). 
\label{eq:kappa6ahigh}
  \end{equation}
We note that for $x=x_{\mu}$, $t=-m^2$. The uncertainty of~\eq{amu6Aspacex} due to the series approximation of $K^{(6a)}$ is estimated to be less than $O(10^{-12})$.

%%%%%%%%%%%%%%%%%%%%%%%%%%
\subsection{Classes $(6b)$ and $(6bll)$}
\label{HVPNNLObblbl}

The contributions of classes $(6b)$ and $(6bll)$ can be calculated similarly to class $(6a)$. Indeed, in the time-like region, $a_{\mu}^{(6b)}$ and $a_{\mu}^{(6bll)}$ can be computed via~\eq{amu6Atime} replacing $ K^{(6a)}$ with $ K^{(6b)}$ and $ K^{(6bll)}$, respectively. For these sixth-order kernel functions, approximate series expansions in the parameters $r=m^2/s$ and $\rho=m_e/m$ were computed in~\cite{Kurz:2014wya}. The highest order expansion terms provided are of $O(\rho^2 r^4)$. Following the same procedure described in Subsection \ref{HVPNNLOa}, we fit these expansions obtaining integral representations analogous to that of~\eq{K6A} with the coefficients $g_i^{(6b)}$, $h_i^{(6b)}$, $j_i^{(6b)}$, $p_i^{(6b)}$ and $g_i^{(6bll)}$, $h_i^{(6bll)}$, $j_i^{(6bll)}$, $p_i^{(6bll)}$ reported in Table~\ref{table:6b} and~\ref{table:6bll}, respectively. The contributions $a_{\mu}^{(6b)}$ and $a_{\mu}^{(6bll)}$ can then be calculated in the space-like region using Eqs.~(\ref{eq:amu6Aspacex}--\ref{eq:kappa6ahigh}), mutatis mutandis. Their estimated uncertainties due to the series approximations are less than $O(10^{-12})$.

%%%%%%%%%%%%%%%%%%%%%%%%%%
\subsection{Class $(6c)$}
\label{HVPNNLOc}

The contribution of class $(6c)$ in the time-like region is given by~\cite{Kurz:2014wya}
\begin{align}
	a_{\mu}^{(6c)} = \,\,  &\frac{\alpha^2}{\pi^4} \int_{s_0}^{\infty} \frac{{\rm d}s}{s} \frac{{\rm d}s'}{s'}  
	\, K^{(6c)}(s/m^2,s'/m^2) \, \times \notag \\
	  & \times {\rm Im} \Pi_{\rm h} (s) \,  {\rm Im} \Pi_{\rm h} (s').
\label{eq:amu6Ctime}
\end{align}
As class $(6c)$ diagrams contain two HVP insertions, the time-like formula~(\ref{eq:amu6Ctime}) for $a_{\mu}^{(6c)}$ requires two dispersive integrations of ${\rm Im} \Pi_{\rm h} (s)$. Asymptotic expansions were provided in Ref.~\cite{Kurz:2014wya} for the function $K^{(6c)}(s/m^2,s'/m^2)$ in the limits $s' \approx s \gg m^2$ and $s' \gg s \gg m^2$, from which an approximation of $K^{(6c)}(s/m^2,s'/m^2)$ valid for all values of $s'$ and $s$ much larger than $m^2$ can be constructed.

In the time-like approach, the number of dispersive integrations of ${\rm Im} \Pi_{\rm h} (s)$ required to calculate the contribution of a diagram to the muon $g$-2 is given by the number of HVP insertions. On the other hand, the required dimension of the space-like integral of $\Delta \alpha_{\rm h}(t)$ (or powers of it) equals the number of photon lines with different momenta containing HVP insertions. To obtain a space-like formula for $a_{\mu}^{(6c)}$, it is therefore convenient to separate the diagrams of class $(6c)$ into the following four subclasses $(6c1)$, $(6c2)$, $(6c3)$, and $(6c4)$ (see Fig.~\ref{fig:NNLOdiagrams}).

The diagrams of subclass $(6c1)$ contain two HVP insertions in the same photon line and no other electron or muon loop. The exact space-like expression for their contribution to the muon $g$-2 is therefore given by the one-dimensional integral
\begin{equation}
	a_{\mu}^{(6c1)}  = \left( \frac{\alpha}{\pi} \right)^2 \! \int_{0}^{1} {\rm d}x \, \lambda^{(4)} (x) \, \left[\Delta \alpha_{\rm h}( t(x) ) \right]^2,
\label{eq:amu6C1spacex}
\end{equation}
where the kernel function is
\begin{equation}
	\lambda^{(4)} (x) \,=\,  \kappa^{(4)} (x)  \,-\, \frac{2\pi}{\alpha} \kappa^{(2)} (x) \, \Delta \alpha_{\mu}^{(2)}( t(x) ),
\label{eq:lambda4}
\end{equation}
$\kappa^{(4)} (x)$ is the exact fourth-order space-like kernel of~\eq{kappa4} and $\kappa^{(2)} (x)$ is the lowest-order one of~\eq{kappa2}. In~\eq{amu6C1spacex}, the use of the subtracted kernel $\lambda^{(4)} (x)$ instead of $\kappa^{(4)} (x)$ guarantees the subtraction of the contribution, induced by $\kappa^{(4)} (x)$, of two diagrams containing two HVP and one muon VP in the same photon line.

The contribution of the three diagrams of subclass $(6c3)$, containing two HVP and one electron VP insertion in the same photon line, can be cast in the exact space-like one-dimensional integral form
\begin{equation}
	a_{\mu}^{(6c3)}  = \frac{3\alpha}{\pi} \int_{0}^{1} {\rm d}x \, \kappa^{(2)} (x)
	\left[\Delta \alpha_{\rm h}( t(x) ) \right]^2 \!
	\Delta \alpha_e^{(2)}( t(x) ).
\label{eq:amu6C3spacex}
\end{equation}
Analogously, the exact contribution of subclass $(6c4)$, comprising three diagrams with two HVP and one muon VP insertion in the same photon line, can be simply obtained replacing $\Delta \alpha_e^{(2)}( t )$ with $\Delta \alpha_{\mu}^{(2)}( t )$ in~\eq{amu6C3spacex}.

Subclass $(6c2)$ consists of diagrams with two HVP insertions in two different photon lines. Contrary to the simple one-dimensional integral form of all the space-like expressions for the contributions to the muon $g$-2 discussed so far, the presence in $(6c2)$ of two different photon lines with HVP insertions requires a double space-like integration. We therefore proceeded in two steps. 
First, we computed the approximate time-like kernel $K^{(6c2)}(s/m^2,s'/m^2)$ for the subclass $(6c2)$. This was obtained by calculating the exact time-like kernels $K^{(6c1)}(s/m^2,s'/m^2)$, $K^{(6c3)}(s/m^2,s'/m^2)$ and $K^{(6c4)}(s/m^2,s'/m^2)$ from the exact space-like expressions of~Eqs.(\ref{eq:amu6C1spacex},\ref{eq:amu6C3spacex}), computing the series expansion of these kernels in the limits $s' \approx s \gg m^2$ and $s' \gg s \gg m^2$, and finally subtracting the obtained results from the $K^{(6c)}(s/m^2,s'/m^2)$ approximation of Ref.~\cite{Kurz:2014wya}.
As a second step, we matched the LO terms of the approximate time-like kernel $K^{(6c2)}(s/m^2,s'/m^2)$ with those of the series expansion of a two-dimensional generating integral representation, generalising to two-dimensions the method used earlier to fit the $K^{(6a)}(s/m^2)$ expansion. Our result for the space-like expression of the contribution of subclass $(6c2)$ to the muon $g$-2 is
\begin{align}
	a_{\mu}^{(6c2)} \, = \,  & \left( \frac{\alpha}{\pi} \right)^2 \! \int_{x_{\mu}}^{1} \! {\rm d}x \int_{x_{\mu}}^{1} \! {\rm d}x'  \,  \bar{\kappa}^{(6c2)} (x,x')  \, \times 						\notag \\
	                              & \times  \Delta \alpha_{\rm h}( t(x) )  \Delta \alpha_{\rm h}( t(x') ),
\label{eq:amu6C2spacex}
\end{align}
where, for $x_{\mu} < \{x,x'\} < 1$,
\begin{equation}
	\bar{\kappa}^{(6c2)} (x,x') \,=\, \frac{2-x}{x^3} \frac{2-x'}{x'^3} \, G^{(6c2)} \!\!  \left(\frac{1-x}{x^2},\frac{1-x'}{x'^2}\right)
\label{eq:kappa6c2high}
  \end{equation}
and
\begin{align}
 	\!\!G^{(6c2)}(\xi,\xi') \, = \,\,   \frac{1}{4 \left(32 \pi^2 -315 \right)} \, & \times  \notag \\
				\times \biggl[ \left(1855-188 \pi^2 \right) 		& \, \frac{\min(\xi,\xi')}{\max(\xi,\xi')^2}  \, + \notag \\
					 + \left(988 \pi^2-9765 \right) 			& \,\frac{\min(\xi,\xi')^2}{\max(\xi,\xi')^3} \, + \notag \\
					 + 24 \left(435 - 44\pi^2 \right) 			& \, \frac{\min(\xi,\xi')^3}{\max(\xi,\xi')^4} \biggr].
\label{eq:G6C2}
\end{align}
This contribution is of $O(10^{-12})$. We note that the limits of integration in~\eq{amu6C2spacex} are $x_{\mu}$ and 1, corresponding to values of $t$ between $-m^2$ and $-\infty$, respectively.

Equation~(\ref{eq:amu6C2spacex}) completes the list of space-like expressions for the contributions of class $(6c)$,
\begin{equation}
	a_{\mu}^{(6c)} = a_{\mu}^{(6c1)} + a_{\mu}^{(6c2)} + a_{\mu}^{(6c3)} + a_{\mu}^{(6c4)}.
\label{eq:amu6Cspacex}
\end{equation}
The uncertainty of~\eq{amu6Cspacex} due to the approximations of subclass $(6c2)$ is less than $O(10^{-12})$.

%%%%%%%%%%%%%%%%%%%%%%%%%%
\subsection{Class $(6d)$}
\label{HVPNNLOd}

The correction due to the single diagram of class $(6d)$ can be written in the time-like form~\cite{Kurz:2014wya}
\begin{align}
	a_{\mu}^{(6d)} \,= \,\,  &\frac{\alpha}{\pi^4} \int_{s_0}^{\infty} \frac{{\rm d}s}{s} \frac{{\rm d}s'}{s'} \frac{{\rm d}s''}{s''}  
	\, K^{(6d)}(s,s',s'') \, \times \notag \\
	  & \times {\rm Im} \Pi_{\rm h} (s) \,  {\rm Im} \Pi_{\rm h} (s') \,  {\rm Im} \Pi_{\rm h} (s'').
\label{eq:amu6Dtime}
\end{align}
The kernel $K^{(6d)}(s,s',s'')$ for the triple hadronic insertion is provided in~\cite{Kurz:2014wya} in integral form. On the other hand, the space-like expression for $a_{\mu}^{(6d)} $ can be cast in the simple exact form~\cite{Jegerlehner:2017gek}
\begin{equation}
	a_{\mu}^{(6d)}  \,=\,  \frac{\alpha}{\pi}  \int_{0}^{1} {\rm d}x \, \kappa^{(2)} (x) \, \left[\Delta \alpha_{\rm h}( t(x) ) \right]^3.
\label{eq:amu6Dspacex}
\end{equation}
We note that three dispersive integrations of ${\rm Im} \Pi_{\rm h} (s)$ are required to compute $a_{\mu}^{(6d)}$ in the time-like approach, whereas the space-like~\eq{amu6Dspacex} involves only a one-dimensional integral. Numerically, $a_{\mu}^{(6d)}$ is very small, of $O(10^{-13})$.

%%%%%%%%%%%%%%%%%%%%%%%%%%%%%%%%%%%%%%%%%%%%%%%%%%%%%%%%%%%%
\section{Conclusions}
\label{Conclusions}

This paper provides simple analytic expressions to calculate the HVP contributions to the muon $g$-2 in the space-like region up to NNLO. These results can be employed in lattice QCD computations of $a_{\mu}^{\rm HVP}$ as well as in determinations based on scattering data, like those expected from the proposed MUonE experiment at CERN. 

After a derivation of the space-like formula for the HVP contribution at LO, obtained using the dispersion relation satisfied by the LO time-like kernel $K^{(2)}(z)$, we presented simple exact analytic expressions to extend the space-like calculation of $a_{\mu}^{\rm HVP}$ to NLO. The shapes of the space-like integrands of the $a_{\mu}^{\rm HVP}({\rm NLO})$ contributions were found to differ significantly from the LO one. In particular, the exact NLO space-like kernel $\kappa^{(4)} (x)$ provides a stronger weight to $\Delta \alpha_{\rm h}(q^2)$ at large negative values of $q^2$ than the LO kernel $\kappa^{(2)} (x)$. These different weights may help to shed light on the present tension between the lattice QCD determination of $a_{\mu}^{\rm HVP}({\rm LO})$ by the BMW collaboration and the time-like data-driven ones.

The approximation to the NLO space-like kernel $\kappa^{(4)} (x)$ obtained by the authors of Ref.~\cite{Chakraborty:2018iyb} induced the largest source of uncertainty of their NLO lattice QCD calculation of $a_{\mu}^{\rm HVP}({\rm NLO})$. This uncertainty, of $O(10\%)$, can be eliminated using the exact expression for $\kappa^{(4)} (x)$ provided in this paper.

The NNLO HVP contribution to the muon $g$-2 is comparable to the final uncertainty expected from the Muon $g$-2 experiment at Fermilab. We presented simple analytic space-like expressions for all the classes of diagrams representing these corrections. For the diagrams composed of one- or two-loop QED vertices and two or more HVP insertions in the same photon line, we obtained exact space-like integral formulas. For the diagrams containing actual three-loop QED vertices, like e.g.\ electron or muon light-by-light graphs, exploiting generating integral representations to fit the large-$s$ approximate series expansions of the time-like kernels provided by Ref.~\cite{Kurz:2014wya}, we found very good approximations to the space-like kernels. The uncertainty of $a_{\mu}^{\rm HVP}({\rm NNLO})$ due to these kernel approximations is estimated to be less than $O(10^{-12})$.

In the space-like approach, the minimum dimension of the space-like integral of $\Delta \alpha_{\rm h}(t)$ (or powers of it) required to calculate the contribution of a diagram to the muon $g$-2 is given by the number of photon lines with different momenta containing HVP insertions. The space-like kernels to compute $a_{\mu}^{\rm HVP}$ at LO and NLO are therefore one-dimensional. The same is true at NNLO, with the notable exception of the class of diagrams with two HVP insertions in two different photon lines. For this class, a two-dimensional kernel is required. Generalising to two dimensions the one-dimensional method used earlier, we derived a good approximate two-dimensional space-like kernel matching the approximate time-like kernel with the series expansion of a two-dimensional generating integral representation. Once again, the uncertainty due to the kernel approximation is less than $O(10^{-12})$.

The calculation of higher-order HVP corrections to the muon $g$-2 requires a precise treatment of the QED radiative corrections to the HVP function. Their leading effect, induced by the emission and reabsorption of a photon by the HVP insertion, is normally incorporated into the time-like approach via the inclusion of final-state radiation corrections in the $R$-ratio. This is a notoriously delicate issue, because of the experimental cuts imposed by the analyses. On the other hand, the fully inclusive measurement of $\Delta \alpha_{\rm h}(t)$ expected from MUonE will naturally include these leading corrections in the space-like approach. 

In conclusion, the results presented in this paper allow to compare, for the first time, time-like and space-like calculations of $a_{\mu}^{\rm HVP}$ at NNLO accuracy. These eagerly anticipated comparisons will strengthen the SM prediction of the muon $g$-2 enhancing its potential to unveil new physics.

%%%%%%%%%%%%%%%%%%%%%%%%%%%%%%%%%%%%%%%%%%%%%%%%%%%%%%%%
~

{\em Acknowledgments} We would like to thank V.~Barigelli, G.~Colangelo, M.~Fael, M.~Hoferichter, A.~Keshavarzi, W.~J.~Marciano, M.~Steinhauser and G.~Venanzoni for useful discussions and correspondence. We are also grateful to all our MUonE colleagues for our stimulating collaboration.
M.~P.\ acknowledges partial support from the EU Horizon 2020 research and innovation programme under the Marie Sklodowska-Curie grant agreements 101006726 (aMUSE) and 789410 (HIDDeN).

%%%%%%%%%%%%%%%%%%%%%%%%%%%%%%%%%%%%%%%%%%%%%%%%%%%%%%%%%%%%
\bibliography{mybibfile}

%111%%%%%%%%%%%%%%%%%%%%%%%%%%%%%%%%%%%%%%%%%%%%%%%%%%%%%%%%%%%
\begin{table*}[]
\renewcommand{\arraystretch}{1.5}
\begin{tabular}{cl}
\hline
 \multicolumn{2}{|p{\textwidth}|}{ \centering{\boldmath{$(6a)$} \unboldmath}}\\ \hline
 \multicolumn{1}{|p{0.49\textwidth}|}{\begin{tabular}[c]{@{}l@{}}$j_0=0$;\\ 
 $j_1=-\frac{3793}{864}$;\\ 
 $j_2=\frac{35087}{21600}$;\\ 
 $j_3= \frac{1592093}{43200}$;\end{tabular}} &  \multicolumn{1}{|p{0.4\textwidth}|}{\begin{tabular}[c]{@{}l@{}}$h_0=-\frac{359}{36}$;\\
 $h_1=\frac{122293}{5184}$;\\
 $h_2=-\frac{43879427}{648000}$;\\ 
 $h_3=\frac{14388407}{48000}$;\end{tabular}}                                                                                                                                                                                                                                                                                                                                                                                                                                                                                                                                                                                                                                                                                                                                                                                                                                                                                                                                                                                                                                                                                                                                                                                                                                                                                                                                                                                                                                                                                                                                                                                                                                                                                                                                                                                                                                                                                                                                                                                                                                                                                \\ \hline
 \multicolumn{2}{|l|}{\begin{tabular}[c]{@{}l@{}}$g_0=\frac{1301}{144}-\frac{19\pi^2}{9}$;\\ 
 $g_1=\frac{441277}{10368}+\pi^2\left(-\frac{355}{648}+\ln{4}\right)+\frac{25 \ \zeta(3)}{2}$;\\
 $g_2=-\frac{5051645167}{38880000}+\pi^2\left(\frac{221411}{32400}-18\ln{2}\right)-\frac{3919\ \zeta(3)}{60}$;\\ 
 $g_3=\frac{14588342017}{38880000}+\pi^2\left(-\frac{2479681}{64800}+112\ln{2}\right)+\frac{3113 \ \zeta(3)}{10}$;\end{tabular}}                                                                                                                                                                                                                                                                                                                                                                                                                                                                                                                                                                                                                                                                                                                                                                                          \\\hline
 \multicolumn{2}{|l|}{\begin{tabular}[c]{@{}l@{}}$p_0=-\frac{1808080780513}{14580000}+\frac{41851\pi^4}{15}+\frac{8432\ln^4{2}}{3}+67456 \ a_4+\frac{2085448 \ \zeta(3)}{15}+$\\
   \ \ \ \ $+\pi^2\left(-\frac{11944163099}{194400}+\frac{272}{3}\left(180-31\ln{2}\right)\ln{2}+\frac{115072 \ \zeta(3)}{3}\right)-\frac{575360 \ \zeta(5)}{3}$;\\ 
   $p_1=\frac{134017456919}{96000}-\frac{4481182\pi^4}{135}-\frac{98420\ln^4{2}}{3}-787360\ a_4+2255200 \ \zeta(5)+$\\ 
   \ \ \ \ $+\pi^2\left(\frac{23549054249}{32400}-201122\ln{2}+\frac{98420\ln^2{2}}{3}-451040 \ \zeta(3)\right)-\frac{57189259 \ \zeta(3)}{36}$;\\ 
   $p_2=-\frac{13069081405453}{3888000}+\frac{330073\pi^4}{4}+80790\ln^4{2}+1938960\ a_4+\frac{77371609 \ \zeta(3)}{20}+$\\
    \ \ \ \ $+\pi^2\left(-\frac{729995599}{405}+6\left(85313-13465\ln{2}\right)\ln{2}+1114360 \ \zeta(3)\right)-5571800 \ \zeta(5)$;\\ 
    $p_3=\frac{1274611832039}{583200}-\frac{986377\pi^4}{18}-53340\ln^4{2}-1280160\ a_4+\frac{11057200 \ \zeta(5)}{3}+$\\ 
    \ \ \ \ $+\pi^2\left(\frac{5809659289}{4860}+420\ln{2}\left(-823+127\ln{2}\right)-\frac{2211440 \ \zeta(3)}{3}\right)-\frac{22833188 \ \zeta(3)}{9}$;\end{tabular}} \\ \hline                                                                                                                                                                                                                                                                                                                                                                                                                                                                                                                                                                                                                                                                                                                                                                                                                                                                                                                                                                                                                                                                                                                                                                                                                        
\end{tabular}
\caption{The coefficients $g_i^{(6a)}$, $h_i^{(6a)}$, $j_i^{(6a)}$, $p_i^{(6a)}$ ($i=0,1,2,3$). The superscript ${(6a)}$ has been dropped for simplicity. In the above coefficients, the Riemann zeta function $\zeta(k) = \sum_{n=1}^\infty 1/n^k$ and $a_4=\sum_{n=1}^{\infty} 1/(2^n n^4)= {\rm Li}_4(1/2)$.}
\label{table:6a}
\end{table*}
%%%%%%%%%%%%%%%%%%%%%%%%%%%%%%%%%%%%%%%%%%%%%%%%%%%%%%%%%%%%

%222%%%%%%%%%%%%%%%%%%%%%%%%%%%%%%%%%%%%%%%%%%%%%%%%%%%%%%%%%%%
\begin{table*}[]
\renewcommand{\arraystretch}{1.5}
\begin{center}
\begin{tabular}{cl}
\hline
 \multicolumn{2}{|p{\textwidth}|}{\centering{\boldmath{$(6b)$} \unboldmath}}                                                                                                                                                                                                                                                                                                                                                                                                                                                                                                                                                                                                                                                                                                                                                                                                                                                                                                                                                                                                                                                                                                                                                                                                                                                                                                                                                                                                                                                                                                                                                                                                                                                                                                                                                     \\ \hline
\multicolumn{1}{|p{0.49\textwidth}|}{\begin{tabular}[c]{@{}l@{}}$j_0=0$;\\ 
$j_1=\frac{11}{27}$;\\ 
$j_2=\frac{41}{120}$;\\ 
$j_3= -\frac{507}{40}$;\end{tabular}} &        \multicolumn{1}{|p{0.4\textwidth}|}{\begin{tabular}[c]{@{}l@{}}$h_0=\frac{65}{54}$;\\ 
$h_1=-\frac{3559}{1296}+\rho^2+\frac{5}{18}\ln{\rho}$;\\ 
$h_2=\frac{3917}{432}-\frac{82 \rho^2}{3}+\frac{61}{10}\ln{\rho}$;\\ 
$h_3=-\frac{4109}{80}+\frac{2211 \rho^2}{10}-\frac{1763}{30}\ln{\rho}$;\end{tabular}}                                                                                                                                                                                                                                                                                                                                                                                                                                                                                                                                                                                                                                                                                                                                                                                                                                                                                                                                                                                                                                                                                                                                                                                                                                                                                                                                                                                                                                                                                                                                                                                                                                                                                                                                                                                                                                                                                                                                                                                                                                                                                                                                                                                                                                                                                                                                                                                                                                                                                                                                                                                                                                                                                                                                                                                                                                                                                                                       \\ \hline
\multicolumn{2}{|l|} {\begin{tabular}[c]{@{}l@{}}$g_0=\frac{1}{108}\left(259-72 \rho^2+276\ln{\rho}\right)$;\\ 
$g_1=-\frac{9215}{1296}+\frac{65\pi^2}{162}-\frac{3\pi^2\rho}{4}+\frac{49\rho^2}{36}+\left(-\frac{301}{54}+8\rho^2\right)\ln{\rho}+\frac{4}{3}\ln^2{\rho}+2 \ \zeta(3)$;\\ $g_2=\frac{501971}{40500}-\frac{113\pi^2}{36}+\frac{270 \pi^2\rho}{36}-\frac{8417\rho^2}{180}+\left(\frac{3479}{900}-44 \rho^2\right)\ln{\rho}-8\ln^2{\rho}-12\ \zeta(3)$;\\ 
$g_3=-\frac{2523823}{324000}+\frac{625\pi^2}{36}-49\pi^2 \rho+\frac{84946 \rho^2}{225}+\left(\frac{987}{50}+200\rho^2\right)\ln{\rho}+\frac{112}{3}\ln^2{\rho}+56\ \zeta(3)$;\end{tabular}}                                                                                                                                                                                                                                                                                                                                                                                                                                                                                                                                                                                                                                                                                                                                                                                                                                                                                                   \\ \hline
 \multicolumn{2}{|l|}{\begin{tabular}[c]{@{}l@{}}$p_0=-\frac{95519053063}{486000}-7275 \pi^2 \rho+\left(-\frac{587150693}{5400}+\frac{75272 \rho^2}{3}+\frac{120800 \pi^{2}}{9}\right)\ln{\rho}+\left(\frac{1135508}{9}+96\rho^2\right)\zeta(3)+$\\ 
 \ \ \ \ $+4720 \ln^2{\rho}+\frac{1067115409 \rho^2}{5400}+ \pi^{2}(\frac{24382331}{810}-\frac{285184}{9}\ln{2})-32\pi^2\rho^2\left(687+\ln{4}\right)$;\\
 $p_1=\frac{279489728279}{121500}+\frac{179283 \pi^{2}\rho}{2}+\left(\frac{2280933773}{1800}-309540\rho^2-\frac{1419328 \pi^{2}}{9}\right) \ln{\rho}-\frac{10}{3}\left(446023+216\rho^2\right)\zeta(3)+$\\
  \ \ \ \  $-\frac{174712}{3} \ln^2{\rho}-\frac{174350167 \rho^2}{75}+\pi^{2}\left(-\frac{143574463}{405}+\frac{3352256 \ln{2}}{9}\right)+\frac{16}{3}\pi^2\rho^2\left(48481+90\ln{2}\right)$;\\$p_2=-\frac{229560199193}{40500}-\frac{912495 \pi^{2}\rho}{4 }+\left(-\frac{1867939691}{600}+788488\rho^2+\frac{1168336 \pi^{2}}{3}\right) \ln{\rho}+\left(\frac{11034553}{3}+1440\rho^2\right)\zeta(3)+$\\ 
  \ \ \ \ $+148348 \ln^2{\rho}+\frac{258653648 \rho^2}{45}+\frac{4}{135} \pi^{2}(29597029-31048560 \ln{2})-\frac{320}{3}\pi^2\rho^2\left(5989+\ln{512}\right)$;\\ $p_3=\frac{72762177677}{19440}+154035 \pi^2 \rho-\frac{7}{108}\left(-31650719+3973440\pi^2+8220240\rho^2\right)\ln{\rho}-\frac{280}{9}\left(78283+27\rho^2\right)\zeta(3)+$\\  
  \ \ \ \ $-100240\ln^2{\rho}-\frac{513692207\rho^2}{135}+\frac{35}{162}\pi^2\left(-2687659+2816064\ln{2}\right)+\frac{140}{3}\pi^2\rho^2\left(9055+\ln{4096}\right)$;\end{tabular}} \\ \hline
\end{tabular}
\end{center}
\caption{The coefficients $g_i^{(6b)}$, $h_i^{(6b)}$, $j_i^{(6b)}$, $p_i^{(6b)}$ ($i=0,1,2,3$). The superscript ${(6b)}$ has been dropped for simplicity. In the above coefficients, $\rho=m_e/m$, the Riemann zeta function $\zeta(k) = \sum_{n=1}^\infty 1/n^k$,  and $a_4=\sum_{n=1}^{\infty} 1/(2^n n^4)= {\rm Li}_4(1/2)$.}
\label{table:6b}
\end{table*}
%%%%%%%%%%%%%%%%%%%%%%%%%%%%%%%%%%%%%%%%%%%%%%%%%%%%%%%%%%%%

%333%%%%%%%%%%%%%%%%%%%%%%%%%%%%%%%%%%%%%%%%%%%%%%%%%%%%%%%%%%%
\begin{table*}[]
\renewcommand{\arraystretch}{1.5}
\begin{center}
\begin{tabular}{cl}
\hline
\multicolumn{2}{|p{\textwidth}|}{\centering{\boldmath{$(6bll)$} \unboldmath}          }                                                                                                                                                                                                                                                                                                                                                                                                                                                                                                                                                                                                                                                                                                                                                                                                                                                                                                                                                                                                                                                                                                                                                                                                                                                                                                                                                                                                                                                                                                                                                                                                                                                                                                                                                                                                                                                                                                                                                                                                                                                                                                                                                                                                                                                                                                                   \\ \hline
\multicolumn{1}{|p{0.49\textwidth}|}{\begin{tabular}[c]{@{}l@{}}$j_0=0$;\\ 
$j_1=\frac{4}{27}-\frac{9 \rho^2}{2}$;\\ 
$j_2=-\frac{41}{48}+\frac{2201 \rho^2}{216} $;\\ 
$j_3= \frac{3037}{900}-\frac{5909 \rho^2}{216}$;\end{tabular} } & \multicolumn{1}{|p{0.4\textwidth}|}{ \begin{tabular}[c]{@{}l@{}}$h_0=-\frac{9}{2}$;\\ 
$h_1=\frac{59}{9}-\frac{275 \rho^2}{36}-18\rho^2\ln{\rho}$;\\ 
$h_2=-\frac{485}{32}+\frac{1351\rho^2}{48}+\frac{659 \rho^2}{18}\ln{\rho}$;\\ 
$h_3=\frac{282617}{6750}-\frac{10481 \rho^2}{108}-\frac{851 \rho^2}{9}\ln{\rho}$;\end{tabular}          }                                                                                                                                                                                                                                                                                                                                                                                                                                                                                                                                                                                                                                                                                                                                                                                                                                                                                                                                                                                                                                                                                                                                                                                                                                                                                                                                                                                                                                                                                                                                                                                                                                                                                                                                                                                                                                                                                                                                                                                                                                                                                                                                                                                                                                                                                                                                                                                                                                                                                                                                                                                                                                                                                                                                                                                                                                                                                                                                                                                                                                                                                                                                                                                                                                                                                                                                                                                                                                                                                                                                                                                                                                                                                                                                                                                                                                                                                                                                                                                              \\ \hline \multicolumn{2}{|l|}{\begin{tabular}[c]{@{}l@{}}$g_0=\frac{43}{8}-4\pi^2 \rho +15\rho^2 +\pi^2\rho^2-18\rho^2\ln{\rho}+6\rho^2\ln^2{\rho}$;\\ 
$g_1=-\frac{73}{81}+\frac{8\pi^2}{81}+\frac{40 \pi^2 \rho}{9}+\frac{2437 \rho^2}{108}+\frac{17 \pi^2 \rho^2}{9}+\frac{607 \rho^2}{18}\ln{\rho}-\frac{20\rho^2}{3}\ln^2{\rho}+\frac{2}{3} \zeta(3)+2\rho^2\zeta(3)$;\\ 
$g_2=-\frac{385}{162}-\frac{41\pi^2}{72}-\frac{28 \pi^2 \rho}{3}-\frac{89873 \rho^2}{5184}-\frac{997 \pi^2\rho^2}{324}-\frac{1961 \rho^2}{72}\ln{\rho}+14\rho^2\ln^2{\rho}-\frac{5}{2}\zeta(3)-\frac{16\rho^2}{3}\zeta(3)$;\\ 
$g_3=\frac{2691761}{202500}+\frac{3037\pi^2}{1350}+24\pi^2\rho+\frac{655429 \rho^2}{97200}+\frac{2359 \pi^2\rho^2}{324}+\frac{6943 \rho^2}{360}\ln{\rho}-36\rho^2\ln^2{\rho}+\frac{42}{5}\zeta(3)+15\rho^2\zeta(3)$;\end{tabular} }                                                                                                                                                                                                                                                                                                                                                                                                                                                                                                                                                                                                                                                                                                                                                                                                                                                                                                                                                                                                                                                                                                                                                                        \\ \hline
\multicolumn{2}{|l|}{ \begin{tabular}[c]{@{}l@{}}$p_0=-\frac{343277101}{45000}-\frac{33156604927\rho^2}{583200}+\pi^2\left(-\frac{615427}{4050}+\frac{6776\rho}{3}+\frac{763121\rho^2}{972}\right)-\frac{4\pi^4}{135}\left(7817+3212\rho^2\right)+$\\ 
\ \ \ \ $+\left(-\frac{7290521}{3240}+\frac{49622\pi^2}{27}-\frac{128\pi^4}{9}\right)\rho^2\ln{\rho}+\left(-3388-\frac{80\pi^2}{3}\right)\rho^2\ln^2{\rho}+$\\ 
\ \ \ \  $+\left(25642+\frac{1515724\rho^2}{27}-128\pi^2\rho^2-160\rho^2\ln{\rho}\right)\zeta(3)-\frac{1280}{3}\rho^2\zeta(5)$;\\ 
$p_1=\frac{89280434843}{972000}+\frac{248834878697\rho^2}{388800}-\frac{1}{324}\pi^2\left(-533001+9110736\rho+3110417\rho^2\right)+\frac{2}{135}\pi^4\left(180247+73530\rho^2\right)+$\\ 
\ \ \ \ $+\left(\frac{11101973}{1080}-\frac{193400\pi^2}{9}+\frac{320\pi^4}{3}\right)\rho^2\ln{\rho}+\frac{2}{3}\left(63269+300\pi^2\right)\rho^2\ln^2{\rho}+$\\
 \ \ \ \  $+\frac{1}{45}\left(-13410977+100\left(-292301+432\pi^2\right)\rho^2+54000\rho^2\ln{\rho}\right)\zeta(3)+3200\rho^2\zeta(5)$;\\ 
 $p_2=-\frac{6209532853}{27000}-\frac{29997466847\rho^2}{19440}+\pi^2\left(-\frac{114521}{30}+71840\rho+\frac{1970140\rho^2}{81}\right)-\frac{4}{9}\pi^4\left(14685+6032\rho^2\right)+$\\ 
 \ \ \ \  $-\frac{1}{54}\left(190613-2847360\pi^2+11520\pi^4\right)\rho^2\ln{\rho}-80\left(1347+5\pi^2\right)\rho^2\ln^2{\rho}+$\\ 
 \ \ \ \ $-\frac{10}{9}\left(-658509+\left(-1431463+1728\pi^2\right)\rho^2+2160\rho^2\ln{\rho}\right)\zeta(3)-6400\rho^2\zeta(5)$;\\ 
 $p_3=\frac{49726331179}{324000}+\frac{7324831423\rho^2}{7290}+\pi^2\left(\frac{3897971}{1620}-\frac{145880\rho}{3}-\frac{3977785\rho^2}{243}\right)+\frac{14}{27}\pi^4\left(8269+3419\rho^2\right)+$\\ 
 \ \ \ \  $+\frac{7}{81}\left(-81551-401520\pi^2+1440\pi^4\right)\rho^2\ln{\rho}+\frac{140}{3}\left(1563+5\pi^2\right)\rho^2\ln^2{\rho}+$\\ 
 \ \ \ \ $+\frac{35}{27}\left(-371889+16\left(-50437+54\pi^2\right)\rho^2+1080\rho^2\ln{\rho}\right)\zeta(3)+\frac{11200}{3}\rho^2\zeta(5)$;\end{tabular}} \\ \hline
\end{tabular}
\end{center}
\caption{The coefficients $g_i^{(6bll)}$, $h_i^{(6bll)}$, $j_i^{(6bll)}$, $p_i^{(6bll)}$ ($i=0,1,2,3$). The superscript ${(6bll)}$ has been dropped for simplicity. In the above coefficients, $\rho=m_e/m$, the Riemann zeta function $\zeta(k) = \sum_{n=1}^\infty 1/n^k$, and $a_4=\sum_{n=1}^{\infty} 1/(2^n n^4)= {\rm Li}_4(1/2)$.}
\label{table:6bll}
\end{table*}
%%%%%%%%%%%%%%%%%%%%%%%%%%%%%%%%%%%%%%%%%%%%%%%%%%%%%%%%%%%%

\end{document}